\documentclass[12pt]{article}
\usepackage{latexsym}

\newcommand{\dd}{\mbox{\rm d}}
\newcommand{\wg}{\wedge}
\newcommand{\gam}{\gamma}
\newcommand{\Gam}{\Gamma}
\newcommand{\dg}{\dagger}
\newcommand{\ddg}{\ddagger}
\newcommand{\tl}{\tilde}

\newcommand{\DD}{\mbox{\rm D}}

\newcommand{\oo}{\over}
\newcommand{\p}{\partial}

\newcommand{\be}{\begin{equation}}
\newcommand{\bear}{\begin{eqnarray}}
\newcommand{\ear}{\end{eqnarray}}
\newcommand{\ee}{\end{equation}}
\newcommand{\lbl}{\label}
\newcommand{\bi}{\bibitem}
\newcommand{\ci}{\cite}

\newcommand{\vs}{\vspace}

\begin{document}

\

\leftline{arXiv: hep-th/0412255}

\baselineskip .7cm 

\rightline{Report IJS/TP-25/04}

\vs{27mm}

\begin{center}

{\LARGE \bf Kaluza-Klein Theory without Extra Dimensions: Curved
Clifford Space}$^\ddg$\footnotetext{$^\dg$A revised version of this paper will
appear in Physics Letters B.}

\vs{3mm}

Matej Pav\v si\v c

Jo\v zef Stefan Institute, Jamova 39,
1000 Ljubljana, Slovenia

e-mail: matej.pavsic@ijs.si

\vs{6mm}

{\bf Abstract}

\end{center}

\vs{2mm}

A theory in which 16-dimensional curved Clifford space ($C$-space) provides
a realization of Kaluza-Klein theory is investigated. No extra dimensions
of spacetime are needed: ``extra dimensions" are in $C$-space.
It is shown that the covariant Dirac equation in $C$-space contains
Yang-Mills fields of the U(1)$\times$ SU(2)$\times$SU(3) group as parts
of the  generalized spin connection of the $C$-space.

\vs{8mm}

\section{Introduction}

There is more to spacetime than usually envisaged in special or general
relativity. Even at the classical level, besides the bosonic coordinates
it is customary to include Grassmann odd coordinates into the game
(see, e.g. \ci{Green}).
They provide a description of spinning degrees of freedom. An alternative
way \ci{Pezzaglia}--\ci{PavsicArena}
of extending spacetime is to consider the corresponding Clifford space
(shortly $C$-space) generated by basis vectors $\gam_\mu$. A point of
$C$-space is described by a set of multivector coordinates
$(s, x^\mu, x^{\mu \nu},...)$ which altogether with the corresponding basis
elements $({\bf 1}, \gam_\mu,\gam_{\mu \nu},...)$ form a
Clifford aggregate or {\it polyvector} $X$. It is well known \ci{Riesz,Teitler}
that the elements of the right or left minimal ideals of Clifford
algebra can be used to represent spinors. Therefore, a coordinate polyvector
$X$ automatically contains spinor as well as bosonic coordinates. In
refs. \ci{PavsicParis,PavsicSaasFee} it was proposed to formulate string
theory in terms of polyvectors, and thus avoid using a higher dimensional
spacetime. Spacetime can be 4-dimensional, whilst the extra degrees of
freedom (``extra dimensions") necessary for consistency of string theory
are in {\it Clifford space}.

In this paper we propose to go even further: 16-dimensional curved
Clifford space can provide a realization of the Kaluza-Klein idea 
\ci{PavsicParis}.
We do not need to assume that spacetime has more than four dimensions.
The ``extra dimensions" are in Clifford space. This approach has seeds
in refs.\ci{CastroPavsicHigher,CastroPavsicConform}, but explicitly it was
formulated in refs.\ci{PavsicParis, CastroUnific}. We will first
investigate some basic aspects of the classical general relativity-like
theory in $C$-space. Then we pass to quantum theory and rewrite the
Dirac-like equation in curved $C$-space and show that the corresponding
generalized spin connection contains Yang-Mills fields describing
fundamental interactions.

Although other authors in a number of very illuminating and penetrating
papers \ci{CliffUnificOthers} have investigated unified models
of fundamental interactions within the framework of Clifford algebra,
they have not fully employed the concept of Clifford space, together with
the $C$-space metric, affine and spin connection,  
polyvector-valued wave
function \ci{PavsicBook}, which all enable to formulate a Kaluza-Klein
like theory in 16-dimensional Clifford space defined over 4-dimensional
spacetime. As far as I know this is a novel approach (see also refs.
\ci{CastroPavsicHigher,CastroPavsicConform,PavsicParis, CastroUnific}).

\section{Clifford space as a generalization of spacetime}.

Since pioneering works by Hestenes \ci{Hestenes}, Clifford algebra
has been extensively investigated (see e.g. refs.\ci{Lounesto}--\ci{Moya}).
Some researchers 
\ci{Pezzaglia}--\ci{PavsicArena} proposed
to replace spacetime with a larger geometric structure which is based
on Clifford algebra. This has led to the concept of Clifford space
(shortly $C$-space).

Suppose we have an $n$-dimensional space $V_n$, not necessarily flat. At
every point $x\in V_n$ we have a flat tangent space, its basis being given
in terms of $n$ orthonormal vectors $\gam_a$, $a=1,2,..., n$ satisfying the
Clifford algebra relations
\be
      \gam_a \cdot \gam_b \equiv {1\oo 2} (\gam_a \gam_b + \gam_b \gam_a)
      = \eta_{ab} {\bf 1}
\lbl{2.1}
\ee
where $\eta_{ab}$ is a pseudo-Euclidean metric whose signature is kept
arbitrary at this stage. The basis vectors $\gam_a$ form a local
basis in $V_n$ and they generate the Clifford algebra ${\cal C}_{M_n}$. The
basis of the latter algebra is given by the set
\be
     \lbrace \gam_A \rbrace = \lbrace {\bf 1}, \gam_{a_1}, \gam_{a_1 a_2},...,
     \gam_{a_1 a_2...a_n} \rbrace \; , \quad a_1 < a_2 < ...<a_r \; , \quad
     r=1,2,...,n
\lbl{2.2}
\ee
where $\gam_{a_1 a_2...a_r} \equiv \gam_{a_1} \wg \gam_{a_2} \wg ... \wg
\gam_{a_r} \equiv {1\oo r!} [\gam_{a_1},\gam_{a_2},...,\gam_{a_r}]$ is the
wedge product.

From a local basis $\lbrace \gam_a \rbrace $ we can switch to a coordinate
basis $\lbrace \gam_\mu \rbrace $ according to the relation
\be
      \gam_\mu = {e_\mu}^a \gam_a
\lbl{2.3}
\ee
where ${e_\mu}^a = \gam_\mu \cdot \gam^a$ is the vielbein field.

The coordinate basis vectors satisfy
\be
     \gam_\mu \cdot \gam_\nu \equiv {1\oo 2} (\gam_\mu \gam_\nu +
     \gam_\nu \gam_\nu ) = g_{\mu \nu}
\lbl{2.4}
\ee
where
$g_{\mu \nu}$ is the metric of $V_n$. We may use $\gam_\mu$ as generators
of Clifford algebra with the basis
\be
   \lbrace \gam_M \rbrace = \lbrace \gam, \gam_{\mu_1},\gam_{\mu_1 \mu_2},...,
   \gam_{\mu_1 ...\mu_n} \rbrace \; , \quad \mu_1 < \mu_2 < ...<\mu_r \; , \quad
     r=1,2,...,n
   \lbl{2.5}
\ee
where $\gam = {\bf 1}$ and
$\gam_{\mu_1 ...\mu_r} \equiv \gam_{\mu_1} \wg \gam_{\mu_2} \wg ... \wg
\gam_{\mu_r}$. Since $\gam_\mu$ and $g_{\mu \nu}$ depend on position, we
have different Clifford algebras ${\cal C}_{V_n}$ at different points
$x \in V_n$. The continuous set of all those algebras over a domain
of $V_n$ forms a manifold ${\cal C}_{V_n}(x)$ which is usually called
Clifford bundle  or Clifford manifold.

{\it In this paper we propose to introduce a more general Clifford manifold}
(see also \ci{PezzagliaSpin,PavsicBook,CastroPavsicHigher}). Let us
start from the flat Clifford space with basis (\ref{2.2}). We then perform
transition to a curved Clifford space with basis $\lbrace \gam_M \rbrace $
by means of the relation
\be
     \gam_M = {e_M}^A \gam_A
\lbl{2.6}
\ee
where ${e_M}^A$ is the fielbein field in $C$-space. The latter relation is
more general than (\ref{2.3}). Explicitly it reads
\bear
    &&\gam = {e_{\bf o}}^{\underline {\bf o}}\, {\bf 1} + 
    {e_{\bf o}}^{a_1} \gam_{a_1} + {e_{\bf o}}^{a_1 a_2} \gam_{a_1 a_2} +  ...
    + {e_{\bf o}}^{a_1 ... a_n} \gam_{a_1 ... a_n} \nonumber \\
    &&\gam_{\mu_1} = {e_{\mu_1}}^{\underline {\bf o}} \, {\bf 1} + 
    {e_{\mu_1}}^{a_1} \gam_{a_1}
    + {e_{\mu_1}}^{a_1 a_2} \gam_{a_1 a_2} + ... +
    {e_{\mu_1}}^{a_1 ... a_n} \gam_{a_1 ... a_n} \nonumber \\
    &&\gam_{\mu_1 \mu_2} = {e_{\mu_1 \mu_2 }}^{\underline {\bf o}}\, 
    {\bf 1} + 
    {e_{\mu_1 \mu_2}}^{a_1} \gam_{a_1}
    + {e_{\mu_1 \mu_2}}^{a_1 a_2} \gam_{a_1 a_2} + ... +
    {e_{\mu_1 \mu_2}}^{a_1 ... a_n} \gam_{a_1 ... a_n} \nonumber \\
    &&\vdots \nonumber \\
    &&\gam_{\mu_1... \mu_n} = {e_{\mu_1 ... \mu_n }}^{\underline {\bf o}} \,
    {\bf 1} + {e_{\mu_1 ...\mu_n}}^{a_1} \gam_{a_1}
    + {e_{\mu_1 ... \mu_n}}^{a_1 a_2} \gam_{a_1 a_2} + ... +
    {e_{\mu_1 ... \mu_n}}^{a_1 ... a_n} \gam_{a_1 ... a_n}
\lbl{2.7}
\ear
where $\gam_{a_1 ... a_r} \equiv \gam_{a_1} \wg ... \wg \gam_{a_r}$,
whilst $\gam_{\mu_1 ... \mu_r}$ are no longer defined as the wedge
product.

From the basis elements $\gam_M$ we can define the metric of $C$-space
according to
\be
      G_{MN} = \gam_M^\ddg * \gam_N
\lbl{2.8}
\ee
Here `$\ddg$' denotes the reversion, that is the operation which reverses
the order of the generators $\gam_a$ (for example, $\gam_{a_1 a_2 a_3}^\ddg
= \gam_{a_3 a_2 a_1}$), whilst `*' denotes the scalar product between
two Clifford numbers $A$ and $B$
\be
     A * B = \langle AB \rangle_0
\lbl{2.9}
\ee

The quantities $\gam_M,~ {e_M}^A,~G_{MN}$ are now assumed to depend on
position in $C$-space which can be parametrized by $C$-space coordinates
vector fields
\be
     X = x^M \gam_M = s \gam + x^{\mu_1} \gam_{\mu_1} +
     x^{\mu_1 \mu_2} \gam_{\mu_1 \mu_2} + ... + 
     x^{\mu_1 ...\mu_n} \gam_{\mu_1 ...\mu_n}
\lbl{2.10}
\ee
In $C$-space the multivector grade is relative to a chosen basis, and a
coordinates transformation in $C$-space in general changes the grade
of $\gam_{\mu_1 ... \mu_r}$. Thus even if an object appears
as a 1-vector with respect to a coordinate basis $\gam_\mu$, it is a
polyvector (a superposition of multivectors) with respect to the
local basis $\gam_a$.

We have thus a curved {\it Clifford space} ($C$-space). A point of
$C$-space is described by coordinates $x^M$. A coordinate basis is $\lbrace
\gam_M \rbrace$, whilst a local (flat) basis is $\lbrace \gam_A \rbrace $.
The tetrad field is given by the scalar product ${e_M}^A = \gam_M^\ddg
* \gam^A$.

The multivector coordinates 
$s,~x^{\mu_1},~x^{\mu_1 \mu_2},...,x^{\mu_1 ... \mu_n}$ provide a description
of oriented $r$-dimensional areas. In refs. \ci{PavsicArena} a physical
interpretation was given, namely that the multivector coordinates 
can be used to describe extended objects, such as closed branes.

\section{On the realization of Kaluza-Klein theory in curved
Clifford space}

The basic idea of Kaluza-Klein theory is that spacetime has more
than four dimensions. The extra dimensions of curved spacetime manifest
as gauge fields describing the fundamental interactions. Instead of
introducing extra dimensions, we can investigate a theory which starts
from 4-dimensional spacetime and then generalize it to curved Clifford
space.

Let us first consider the equation of geodesic in curved $C$-space.
We can envisage that physical objects are described
in terms of $x^M = (s, x^\mu, x^{\mu \nu},...)$. The
first straightforward possibility is to introduce a single
parameter $\tau$ and consider a mapping $\tau \rightarrow x^M = X^M (\tau)$
where $X^M (\tau)$ are 16 embedding functions that
describe a worldline in $C$-space. From the point of view of
$C$-space, $X^M (\tau)$ describe a worldline of a ``point
particle": at every value of $\tau$ we have a {\it point} in
$C$-space. But from the perspective of the underlying
4-dimensional spacetime, $X^M (\tau)$ describe an extended
object, sampled by the center of mass coordinates
$X^\mu (\tau)$ and the coordinates
$X^{\mu_1 \mu_2}(\tau),..., X^{\mu_1 \mu_2 \mu_3 \mu_4} (\tau)$.
They are a generalization of the center of mass coordinates in the sense
that they provide information about the object's 2-vector, 3-vector, and
4-vector extension and orientation\footnote{A systematic and detailed
treatment is in ref. \ci{PavsicArena}.}.

The dynamics of such an object is determined by the action
\bear
     &&I[X] = \int \dd \tau \, ({\dot X}^M {\dot X}^N G_{MN} )^{1/2}
     \lbl{3.1} \\
     &&\delta X^M \, : \quad {1\oo {\sqrt{{\dot X}^2}}} \,
     {\dd \oo {\dd \tau}} \left ( {{\dot X}^M\oo \sqrt{{\dot X}^2} } \right )
     + \Gam_{JK}^M {{{\dot X}^J {\dot X}^K}\oo {\dot X}^2 } = 0
\lbl{3.2}
\ear
Here ${\dot X}^J \equiv \dd X^J/\dd \tau$ is the derivative with respect to an
arbitrary monotonically increasing parameter $\tau$, and $\Gam_{JK}^M$
is the connection defined according to\footnote{For more details see
refs. \ci{Hestenes,PavsicBook, CastroPavsicHigher}}
\be
    \p_M \gam_N = \Gam_{MN}^K \gam_K
\lbl{3.3}
\ee
The above relation is a generalization \ci{CastroPavsicHigher} of the well 
known relation \ci{Hestenes}.

When the derivative $\p_M$ acts on a polyvector $A=A^N \gam_N$ we
obtain the covariant derivative $\DD_M$ acting on the components $A^N$:
\be
    \p_M (A^N \gam_N) = \p_M A^N \gam_N + A^N \p_M \gam_N =
    (\p_M A^N + \Gam_{MK}^N A^K) \gam_N \equiv \DD_M A^N \, \gam_N
\lbl{3.4}
\ee
Here the $A^N$ are {\it scalar} components of $A$, and $\p_M A^N$ is just
the ordinary partial derivative with respect to $X^M$:
\be
    \p_M \equiv \left ( {\p \oo {\p s}},~{\p \oo {\p x^{\mu_1}}},~
    {\p\oo {\p x^{\mu_1 \mu_2}}}, ~{\p \oo {\p x^{\mu_1 ... \mu_n}}} \right )
\lbl{3.5}
\ee

The derivative $\p_M$ behaves as a partial derivative when acting on scalars,
and it defines a connection when acting on a basis
$\lbrace \gam_M \rbrace$. It has turned out very practical\footnote{
Especially when doing long calculation (which is usually the job of a
theoretical physicist) it is much easier and quicker to write $\p_M$
than $\Box_M, ~\nabla_M, ~D_{\gam_M}, ~\nabla_{\gam_M}$ which  all are
symbols used in the literature.}
to use
the easily writable symbol $\p_M$ which ---when acting on a polyvector--- 
cannot be confused with partial derivative.

When inspected from the 4-dimensional spacetime, the equation of geodesic
(\ref{3.2}) contains besides the usual gravitation also other
interactions. They are encoded in the metric components $G_{MN}$ of
$C$-space. Gravity is related to the components $G_{\mu \nu},~\mu,\nu
=0,1,2,3$, while gauge fields are related to the components $G_{\mu {\bar M}}$,
where the index ${\bar M} \neq \nu$ assumes 12 possible values, excluding
the four values $\nu = 0,1,2,3$. In addition, there are also
interactions due to the components $G_{{\bar M} {\bar N}}$, but they
have not the property of the ordinary Yang-Mills fields.

If we now consider the known fundamental interactions of the standard model
we see that besides gravity we have 1 photon described by the abelian
gauge field $A_\mu$, 3 weak gauge bosons described by gauge fields
$W_\mu^a$, $a=1,2,3$,  and 8 gluons described $A_\mu^c$, $c=1,2,...,8$.
Altogether there are 12 gauge fields.

Interestingly, the number of mixed components $G_{\mu {\bar M}} =
(G_{\mu [{\bf o}]},~ G_{\mu[\alpha \beta]},~G_{\mu[\alpha \beta \rho]},~
G_{\mu [\alpha \beta \rho \sigma]})$ of the $C$-space metric tensor
$G_{MN}$ coincides with the number of gauge fields in the standard
model\footnote{The numbers of the independent indices $[{\bf o}],~
[\alpha \beta],~[\alpha \beta \rho],~[\alpha \beta \rho \sigma]$ are
respectively 1,6,4,1 which sums to 12.}. For fixed $\mu$, there are 12
mixed components of $G_{\mu {\bar M}}$ and 12 gauge fields $A_\mu,~
W_\mu^a, A_\mu^c$. This coincidence is fascinating and it may indicate
that the known interactions are incorporated in curved Clifford space.

Good features of $C$-space are the following:
\begin{description}
 
  \item[\ \ (i)] We do not need to introduce extra dimensions of spacetime.
  We stay with 4-dimensional spacetime $V_4$, and yet we can proceed
  \` a la Kaluza-Klein. The extra degrees of freedom are in $C$-space,
  generated by a basis of $V_4$.
  
  \item[\ (ii)] We do not need to compactify the extra ``dimensions". The
  extra dimensions of $C$-space, namely $s,~x^{\mu \nu},~
  x^{\mu \nu \rho},~x^{\mu \nu \rho \sigma}$ are not just like the ordinary
  dimensions of spacetime considered in the usual Kaluza-Klein theories.
  The coordinates $x^{\mu \nu},~x^{\mu \nu \rho},~x^{\mu \nu \rho \sigma}$
  are related to oriented $r$-surfaces, $r=2,3,4$, by which we sample
  extended objects. Those degrees of freedom are in principle not
  hidden from our direct observation, therefore we do not need to
  compactify such ``internal" space.
  
  \item[(iii)] The number of the mixed metric components $G_{\mu {\bar M}}$
  (for fixed $\mu$) is 12, precisely the same as the number of gauge fields
  in the standard model.
  
\end{description}
  
\section{The Dirac equation in curved $C$-space}

\subsection{Spinors as members of left ideals}

How precisely the curved $C$-space is related to Yang-Mills gauge fields can be
demonstrated by considering a generalization of the Dirac equation to
curved $C$-space.

Let $\Phi (X)$ be a polyvector valued field over coordinates polyvector
field $X= x^M \gam_M$:
\be
    \Phi = \phi^A \gam_A
\lbl{4.1}
\ee
where $\gam_A,~A=1,2,...,16$, is a local (flat) basis of $C$-space
(see eq.(\ref{2.2})) and $\phi^A$  the projections (components) of
$\Phi$ onto the basis $\lbrace \gam_A \rbrace$. We will suppose that in
general $\phi^A$ are complex-valued scalar quantities.

Instead of the basis $\lbrace  \gam_A \rbrace$ one can consider another
basis, which is obtained after multiplying $\gam_A$ by 4 independent
primitive idempotents \ci{Teitler}
\be
    P_i = {1\oo 4} ({\bf 1} + a_i \gam_A + b_i \gam_B + c_i \gam_C) \; , \quad
   i=1,2,3,4
\lbl{4.2}
\ee
such that
\be
    P_i = {1\oo 4} ({\bf 1} +a_i \gam_A)({\bf 1}+b_i \gam_B) \; , 
    \quad \gam_A \gam_B
   = \gam_C \; , \quad c_i = a_i b_i
\lbl{4.3}
\ee
Here $a_i,~b_i,~c_i$ are complex numbers chosen
so that $P_i^2 = P_i$. For explicit and systematic
construction see \ci{Teitler, Mankoc}.

By means of $P_i$ we can form minimal ideals of Clifford algebra. A
basis of left (right) minimal ideal is obtained by taking one of
$P_i$ and multiply it from the left (right) with all 16 elements
$\gam_A$ of the algebra:
\be
   \gam_A P_i \in {\cal I}_i^L \; , \qquad P_i \gam_A \in {\cal I}_i^R
\lbl{4.4}
\ee
Here ${\cal I}_i^L$ and ${\cal I}_i^R$, $i=1,2,3,4$ are four independent
minimal left and right ideals, respectively. For a fixed $i$ there are
16 elements $P_i \gam_A$, but only 4 amongst
them are different, the remaining elements are just
repetition of those 4 different elements.

Let us denote those different elements $\xi_{\alpha i}$, $\alpha=
1,2,3,4$. They form a basis of the $i$-th left ideal. Every Clifford
number can be expanded either in terms of $\gam_A =
({\bf 1},\gam_{a_1}, \gam_{a_1 a_2}, \gam_{a_1 a_2 a_3},
\gam_{a_1 a_2 a_3 a_4})$ or in terms of $\xi_{\alpha i} =
(\xi_{\alpha 1},~\xi_{\alpha 2},~\xi_{\alpha 3},~\xi_{\alpha 4})$:
\be
   \Phi = \phi^A \gam_A = \Psi = \psi^{\alpha i} \xi_{\alpha i} =
\psi^{\tilde A} \xi_{\tilde A}
\lbl{4.5}
\ee
In the last step we introduced a single spinor index ${\tilde A}$
which runs over all 16 basis elements that span 4 independent
left minimal ideals. Explicitly, eq. (\ref{4.5}) reads
\be
   \Psi = \psi^{\tilde A} \xi_{\tilde A} = \psi^{\alpha 1} \xi_{\alpha 1}
  + \psi^{\alpha 2} \xi_{\alpha 2} + \psi^{\alpha 3} \xi_{\alpha 3} +
\psi^{\alpha 4} \xi_{\alpha 4}
\lbl{4.6}
\ee
Eq.(\ref{4.5}) or (\ref{4.6}) represents a direct sum of four independent
4-component spinors, each living in a different left ideal ${\cal I}_i^L$.

In ref. \ci{PavsicBook} it was proposed\footnote{See also  illuminating
works in refs.\ci{PezzagliaSpin,Greider}.} that the polyvector valued wave function
satisfies the Dirac equation in $C$-space:
\be
     \p \Psi \equiv \gam^M \p_M \Psi = 0
\lbl{4.7}
\ee
The derivative $\p_M$ is the same derivative introduced in
eqs. (\ref{3.3}) and (\ref{3.4}). Now it acts on the object
$\Psi$ which is expanded in terms of the 16 basis elements
$\xi_{\tilde A}$, which in turn can be written as a superposition
of basis elements $\gam_A$ of Clifford algebra. The action of
$\p_M$ on the spinor basis elements $\xi_{\tilde A}$ gives
the {\it spin connection}:
\be
    \p_M \xi_{\tilde A} = {{\Gam_M}^{\tilde B}}_{\tl A} \xi_{\tl B}
\lbl{4.8}
\ee
Using the expansion (\ref{4.6}) and eq.\,(\ref{4.8}) we find
\be
    \gam^M \p_M (\psi^{\tl A} \xi_{\tl A}) = \gam^M (\p_M \psi^{\tl A}
  + {{\Gam_M}^{\tl A}}_{\tl B} \psi^{\tl B}) \xi_{\tl A} = 0
\lbl{4.9}
\ee
This is just a generalization of the ordinary Dirac equation in
curved spacetime. Instead of curved spacetime, spin connection and
the Dirac spinor, we have now curved Clifford space, generalized
spin connection and the generalized spinor $\psi^{\tl A}$ which 
incorporates 4 independent Dirac spinors, as indicated in eq. (\ref{4.6}).

We may now use the relations
\be
      {\xi^{\tl A}}^\ddg * \xi_{\tl B} \equiv \langle {\xi^{\tl A}}^\ddg
   \xi_{\tl B} \rangle_S = {\delta^{\tl A}}_{\tl B}
\lbl{4.10}
\ee
and
\be
    \langle {\xi^{\tl C}}^\ddg \gam^M \xi_{\tl A} \rangle_S =
   {(\gam^M)^{\tl C}}_{\tl A}
\lbl{4.10a}
\ee
where the operation $\langle ~~ \rangle_S \equiv {\rm Tr}
\langle~~ \rangle_0$ takes the scalar part of the expression
and then performs the trace. We normalize $\xi_{\tl A}$
so that (\ref{4.10}) is fulfilled. By means of (\ref{4.10})
we can project eq.(\ref{4.9}) onto its component form
\be
     {{(\gam^M)}^{\tl C}}_{\tl A} (\p_M \psi^{\tl A} + 
  {{\Gam_M}^{\tl A}}_{\tl B} \psi^{\tl B} ) = 0
\lbl{4.11}
\ee
The spinor indices ${\tl A},~{\tl B}$ can be omitted and eq.
(\ref{4.11}) written simply as
\be
    \gam^M (\p_M  + \Gam_M) \psi = 0
\lbl{4.12}
\ee

{\it We see that in the geometric form of the Dirac equation (\ref{4.7})
spin connection is automatically present through the action
of the derivative $\p_M$ on a polyvector $\Psi$ written as a
superposition of basis spinors $\xi_{\tl A}$.} The reader has to
be careful (i) not to confuse our symbol $\p_M$ (when acting on a polyvector)
with a partial derivative, (ii) not to miss the fact that $\Psi$ in
eq. (\ref{4.7})
is a Clifford algebra valued object, not just a component
spinor, and (iii) not hastily think that eq. (\ref{4.7}) lacks
covariance.

\subsection{Yang-Mills gauge fields as spin connection in $C$-space}

Let us define generators of the transformations (i.e., local rotations
in $C$-space) according to
\be
    \Sigma_{AB} = \left\{  \begin{array}{ll}
      {1\oo 2} [\gam_A,\gam_B ]\; , & {\rm if} ~A \neq 
      {\bf \underline o},~B \neq {\bf o} \\
      {\bf 1} \gam_B \; , & {\rm if} ~A = {\bf \underline o}
      \end{array}
      \right.
\lbl{4.32}
\ee
We also have $\Sigma_{AB} = {f_{AB}}^C \gam_C$, where ${f_{AB}}^C$
are constants.

A generic transformation in $C$-space which maps a polyvector
$\Psi$ into another polyvector $\Psi'$ is given by
\be
   \Psi' = R \Psi S
\lbl{4.13}
\ee
where 
\be
     R = {\rm e}^{{1\oo 4} \Sigma_{AB} \alpha^{AB}}
     = {\rm e}^{\gam_A \alpha^A}
     \quad {\rm and}
     \quad S = {\rm e}^{{1\oo 4} \Sigma_{AB} \beta^{AB}} 
     = {\rm e}^{\gam_A \beta^A}
\lbl{ 4.13a}
\ee
Here $\alpha^{AB}$ and $\beta^{AB}$, or equivalently 
$\alpha^A = {f_{CD}}^A \alpha^{CD}$ and $\beta^A = {f_{CD}}^A \beta^{CD}$,
are parameters of the transformation.

In general, eq.\,(\ref{4.13}) allows for the transformation which maps a
basis element $\gam_A$ into a mixture of basis elements. In particular, we
have the following three interesting cases:

      (i) $\alpha^{AB} \neq 0, ~~\beta^{AB} = - \alpha^{AB}$. Then we have
\be
    \Psi' = R \Psi R^{-1}
\lbl{4.14}
\ee
This is the transformation which preserves the structure of Clifford algebra,
i.e., it maps the basis elements $ \gam_A$ into another basia element 
$\gam_{A'}$ of the same Clifford algebra.

    (ii) $\alpha^{AB} \neq 0, ~~\beta^{AB} = 0$. Then we have
\be
     \Psi' = R \Psi
\lbl{4.15}
\ee
This is the transformation which maps a basis spinor $\xi_{\alpha i}$ into
another basis spinor $\xi'_{\alpha i}$ belonging to the same left ideal:
\be
    \xi_{\alpha i} \subset {\cal I}_i^L  \rightarrow \xi'_{\alpha i} =
    R \xi_{\alpha i}
   \subset {\cal I}_i^L
\lbl{4.16}
\ee

   (iii) $\alpha^{AB} = 0, ~~\beta^{AB} \neq 0$. Then
\be
    \Psi' = \Psi S
\lbl{4.17}
\ee
This is the transformation that maps right ideal into the right ideals:
\be
    \xi_{\alpha i} \subset {\cal I}_i^R  \rightarrow \xi'_{\alpha i} =  
    \xi_{\alpha i} S
   \subset {\cal I}_i^R
\lbl{4.18}
\ee

In general, for the transformation (\ref{4.13}) we have
\be
    \Psi' = \psi^{\tl A} R \xi_{\tl A} S = \psi^{\tl A} {U_{\tl A}}^{\tl B} 
    \xi_{\tl B}
  = \psi'^{\tl A} \xi_{\tl A}
\lbl{4.19}
\ee
where
\be
    \psi'^{\tl A} = {U^{\tl A}}_{\tl B} \psi^{\tl B}
\lbl{4.20}
\ee
This transformation, in general, mixes right and left ideals. Eq.\,(\ref{4.20})
can be considered as matrix equation in the space spanned by the generalized
spinor indices ${\tl A},~{\tl B}$:
\be
       \psi' = U \psi
\lbl{4.21}
\ee
where $U$ is a $16 \times 16$ matrix, whilst $\psi$ and $\psi'$ are columns
with 16 elements. From (\ref{4.19}),(\ref{4.20}) it follows that
$U = {\hat R} \otimes {\hat S}^{\rm T}$, where ${\hat R}$ and ${\hat S}$ are
$4 \times 4$ matrices representing the Clifford numbers $R$ and $S$.
That is, $U$ is the direct product of ${\hat R}$ and the transpose 
${\hat S}^{\rm T}$
of ${\hat S}$, and it belongs, in general, to the group $GL(4,C) \times GL(4,C)$.
The group is local, because the basis elements $\gam_A$ entering the
definition (\ref{4.14}) depend on position $X$ in $C$-space according
to the relation analogous to (\ref{3.3}), and also the group parameters
$\alpha^A,~\beta^A$ in general depend on $X$.

We now require that the $C$-space Dirac equation is invariant under
the transformations (\ref{4.13}),\,(\ref{4.21}):
\be
    \p' \Psi' = \p \Psi = \gam^M \p_M (\psi^{\tl A} \xi_{\tl A}) 
\lbl{4.23}
\ee
After using eq.(\ref{4.8}) we then find\footnote{More details will be
provided elsewhere.}
\be
     {\Gamma'_{M {\tl A}}}^{\tl B} = {U_{\tl D}}^{\tl B}
     {U^{\tl C}}_{\tl A} {\Gamma_{M {\tl C}}}^{\tl D} +
      \p_M {U^{\tl D}}_{\tl A} {U_{\tl D}}^{\tl B}
\lbl{4.23a}
\ee
This is the transformation for the generalized spin connection (i.e., the
connection in $C$-space). In matrix notation\footnote{The objects
are considered as matrices in the generalized spinor indices ${\tl A},~
{\tl B},~{\tl C},~{\tl D}$.} this reads
\be
    \Gamma'_M = U \Gamma_M U^{-1} + U \p_M U^{-1}
\lbl{4.24}
\ee

We see that $\Gamma_M$ transforms as a non abelian gauge field.
The most general gauge group here is\footnote{The group GL(4,C) is subjected
to further restrictions resulting from the requirement that the transformations
(\ref{4.13}) should leave the quadratic form $\Psi^{\ddg} * \Psi$ invariant.
So we have $\psi'^\ddg * \Psi' = \langle \psi'^\ddg \Psi' \rangle_S =
\langle S^\ddg \Psi'^\ddg R^\ddg R \Psi S \rangle_S = 
\langle \Psi^\ddg \Psi \rangle_S = \Psi^\ddg * \Psi$, provided that
$R^\ddg R = 1$ and $S^\ddg S = 1$. Explicitly, the quadratic form 
reads $\Psi^\ddg * \Psi = \psi^{* \tl A} \psi^{\tl B} z_{{\tl A}{\tl B}}$,
where $z_{{\tl A}{\tl B}}= \xi_{\tl A}^\ddg * \xi_{\tl B}$ is the
spinor metric.}
 GL(4,C) $\times$ GL(4,C).
As  subgroups it contains for instance SL(2,C) and U(1) $\times$
SU(2) $\times$ SU(3). The former group describes the Lorentz
transformations in Minkowski space (which is a subspace of $C$-space),
whilst the latter group coincides with the gauge group of the standard
model which describes electroweak and strong interaction.  We have
thus demonstrated that the generally covariant Dirac equation in
16-dimensional curved $C$-space contains the coupling of spinor
fields $\psi^{\tl A}$ with non abelian gauge fields
${{\Gam_M}^{\tl A}}_{\tl B}$ which
altogether form the spin connection in $C$-space.

Whether this indeed provides a description of the standard model remains to
be fully investigated. But there is further evidence in favor of the above
hypothesis in the fact that a polyvector field $\Psi = \psi^{\tl A} \xi_{\tl A}$
has 16 complex components. Altogether it has 32 real components. This
number matches, for one generation, the number of independent states for
spin, weak isospin and color (together with the corresponding antiparticle
states) in the standard model . A complex polyvector
field $\Psi$ has thus enough degrees of freedom to form a representation
of the group GL(4,C)$\times$ GL(4,C) which contains the Lorentz group
SL(2,C) and the group of the standard model 
U(1)$\times$SU(2)$\times$SU(3). The generators of the group are given
by $\Sigma_{AB}$ defined in eq. (\ref{4.32}).

In the special case of free fields, the $C$-space Dirac equation (\ref{4.7})
decouples into the set of four independent generalized Dirac equations for four
independent 4-component spinors, each living in a different minimal left
ideal:
\be
     i \gam^M \p_M \psi^{\alpha i} = 0 \; , \quad i = 1,2,3,4
\lbl{4.25}
\ee
or explicitly,
\be
    i (\gam^{\bf o} \p_{\bf o} + \gam^\mu \p_\mu + \gam^{\mu \nu} \p_{\mu \nu}
    + \gam^{\mu \nu \rho} \p_{\mu \nu \rho} + \gam^{\mu \nu \rho \sigma}
    \p_{\mu \nu \rho \sigma}) \psi^{\alpha i} = 0
\lbl{4.26}
\ee
A particular solution is
\be
   \psi^{\alpha i} = u^{\alpha i} {\rm e}^{i P_M X^M}
\lbl{4.27}
\ee         
where $u^{\alpha i}$ satisfies
\be
   (\gam^{\bf o} p_{\bf o} + \gam^\mu p_\mu + \gam^{\mu \nu} p_{\mu \nu}
    + \gam^{\mu \nu \rho} p_{\mu \nu \rho} + \gam^{\mu \nu \rho \sigma}
    p_{\mu \nu \rho \sigma})u^{\alpha i} (p_{\bf o}, p_\mu, p_{\mu \nu},
    p_{\mu \nu \rho}, p_{\mu \nu \rho \sigma}) = 0
\lbl{4.28}
\ee
A spinor $\psi^{\alpha i}$ incorporates, besides the linear momentum
excitations, also the area and volume modes, determined by
$ p_{\mu \nu}, p_{\mu \nu \rho}, p_{\mu \nu \rho \sigma} $. Those extra modes
take into account the extended nature of the object. For a nice description
of this latter concept on an example of the quenched minisuperspace
propagator for $p$-branes see ref.\,\ci{AuriliaFuzzy}.

However, in the interactive case (i.e., in curved
$C$-space), we have the set of coupled equations (\ref{4.12}) in which there
occurs the $C$-space spin connection $\Gam_M$. Using eq.\,(\ref{4.8}) we can
calculate the curvature according to
\be
    [\p_M,\p_N] \xi_{\tl A} = {{{R_{MN}}^{\tl B}}}_{\tl A} \, \xi_{\tl B}
\lbl{4.30}
\ee
where
\be
    {{{R_{MN}}^{\tl B}}}_{\tl A} = \p_M {{\Gam_N}^{\tl B}}_{\tl A} -
    \p_N {{\Gam_M}^{\tl B}}_{\tl A} + 
    {{\Gam_M}^{\tl B}}_{\tl C} {{\Gam_N}^{\tl C}}_{\tl A} -
    {{\Gam_N}^{\tl B}}_{\tl C} {{\Gam_M}^{\tl C}}_{\tl A}
\lbl{4.31}
\ee
This is the relation for the Yang-Mills field strength. From the curvature
we can form the invariant expressions, for instance
${R_{MN}}^{{\tl A} {\tl B}}
({\gam^M}^\ddg * \xi_{\tl A})({\gam^N}^\ddg * \xi_{\tl B})$ and
${R_{MN}}^{{\tl A} {\tl B}}{R^{MN}}_{{\tl A} {\tl B}}$ which can be used in
the action as the kinetic term for the fields ${{\Gam_M}^{\tl A}}_{\tl B}$.

Using eq.(\ref{4.32}) we can express the
spin connection in terms of the generators\footnote{We now omit the indices
${\tl A},~{\tl B}$.}:
\be
   \Gam_N = {1\oo 4}  {\Omega^{AB}}_N \Sigma_{AB} = {A_N}^A \gam_A \; , \quad
   {A_N}^A = {1\oo 4} {\Omega^{CD}}_N {f_{CD}}^A
\lbl{4.33}
\ee
Inserting (\ref{4.33}) into (\ref{4.31}) we obtain
\be
     F_{MN}^A = \p_M {A_N}^A - \p_N {A_M}^A + {A_M}^B {A_N}^C {C_{BC}}^A
\lbl{4.34}
\ee
where ${C_{BC}}^A$ are the structure constants of the Clifford algebra:
$[\gam_A,\gam_B] = {C_{AB}}^C \gam_C$.

The $C$-space Dirac equation (\ref{4.12}) can be split according to
\be
   \left [ \gam^{\mu} (\p_\mu + \Gam_\mu ) + \gam^{\bar M} (\p_{\bar M}
   + \Gam_{\bar M}) \right ] \psi = 0
\lbl{4.35}
\ee
where $M=(\mu, {\bar M})$, and ${\bar M}$ assumes all the values
except $M= \mu = 0,1,2,3$.

From eq.(\ref{4.33}) we read that the gauge field $\Gam_M$ contains:

\ \ \ (i) {\it The spin connection of the 4-dimensional gravity}
$\Gam_\mu^{(4)} = {1\oo 8} {\Omega^{ab}}_\mu [\gam_a,\gam_b]$.

\ \ (ii) {The Yang-Mills fields} ${A_\mu}^{\bar A} \gam_{\bar A}$, 
where we have split the local index according to $A=(a,{\bar A})$.
For ${\bar A} = {\bf o}$ (i.e., for the scalar) the latter gauge field is just
that of U(1) group.

\ \ (iii) {\it The antisymmetric potentials} $A_{\mu \nu},~A_{\mu \nu \rho},
~A_{\mu \nu \rho \sigma}$, if we take indices 
$A={\bf o}$ (scalar) and $M= \mu \nu,~ \mu \nu \rho,~\mu \nu \rho \sigma$.

We see that the $C$-space spin connection contains all physically
interesting fields, including the antisymmetric gauge fields which
occur in string and brane theories.

A caution is in order here. From eq.\,(\ref{4.33}) it appears that there
are only 16 independent generators $\gam_A$ in terms of which a gauge field
is expressed. But inspecting the generators (\ref{4.32}) we see that there are
more than 16 different rotations in $C$-space. However, some of them,
although being physically different transformations, turn out to be
mathematically described by the same objects. For instance, the generators
$\Sigma_{ab} = {1\oo 2} [\gam_a,\gam_b]$ and $\Sigma_{{\tl a}{\tl b}}
={1\oo 2} [\gam_5 \gam_a, \gam_5 \gam_b] = {1\oo 2} [\gam_a,\gam_b]$
are equal, although the corresponding transformations, i.e., a rotation
in the subspace $M_4$ and the rotation in the dual space ${\tl M}_4$ are
in principle independent. Similarly we have $\Sigma_{1 {\tl 1}} =
{1\oo 2} [\gam_1, \gam_5 \gam_1] = \gam_5 = \Sigma_{{\bf o}{\tl {\bf o}}}
= {\bf 1} \gam_5$. Such degeneracy of the transformations
is removed by the fact that the transformation
can act on the spinor polyvector $\Psi$  either from the left or
from the right (according to (\ref{4.13})). 

Returning to the Dirac equation (\ref{4.35}) we see that besides the
part having essentially the same form as the ordinary Dirac equation in the
presence of minimally coupled 4-dimensional spin connection and 
Yang-Mills fields, there is also an extra term which can have the role of mass
if $\psi$ is an eigenstate of the operator $\gam^{\bar M} (\p_{\bar M}
+ \Gam_{\bar M})$. Since the metric signature of $C$-space is \ci{PavsicParis}
$(8+,8-)$ and the signature of the ``internal" space is $(7+,5-)$, the mass
is not necessarily of the order of the  Planck mass; it can be small due
to cancellations of the positive and negative contributions.

\section{Conclusion}

The theory that we pursue here\footnote{See also references 
\ci{CastroPavsicHigher,CastroPavsicConform,PavsicParis,CastroUnific}.}
seems to be a promising candidate for the unification of fundamental
interactions. It does not require inclusion of extra dimensions. Instead,
it employs the degrees of freedom incorporated in the 16-dimensional
Clifford space of the 4-dimensional spacetime. A curved Clifford space
provides an interesting realization of Kaluza-Klein theory without the
necessity of enlarging the dimensionality of spacetime. Such a fresh approach
to unification which takes into account the ideas from various fashionable
theories, e.g., Kaluza-Klein theory, Clifford algebra, string and brane
theory (branes sampled by Clifford numbers), is in my opinion very
promising and deserves further more detailed investigation.

\centerline{Acknowledgement}

This work was supported by the Ministry of
High Educution, Science and Technology of Slovenia

\end{document}